
\magnification=\magstep1
\baselineskip=14pt
\def\nms{\mathsurround=0pt}
\def\overapprox#1#2{\lower2pt\vbox{\baselineskip0pt\lineskip - 1pt
    \ialign{$\nms#1\hfil##\hfil$\crcr#2\crcr\approx\crcr}}}
\def\gtsim{\mathrel{\mathpalette\oversim>}} 
\def\ltsim{\mathrel{\mathpalette\oversim<}} 
\def\oversim#1#2{\lower2pt\vbox{\baselineskip0pt\lineskip 1pt
    \ialign{$\nms#1\hfil##\hfil$\crcr#2\crcr\sim\crcr}}}

\def \Po {\Phi _{\circ}}
\def \ft {\overline F_{t}}
\def \fbk {\overline F_{BK}}
\centerline{\bf SELF-SIMILARITY OF FRICTION LAWS}
\vskip 3cm
\centerline {Maria de Sousa Vieira and Hans J. Herrmann}
\bigskip
\centerline{\sl Hochstleistungsrechenzentrum Supercomputing Center,
Kernforschungsanlage}
\centerline{\sl D-52425 J\"ulich }
\centerline{\sl Germany}
\vskip 3cm
\centerline {ABSTRACT}
\bigskip
{\sl The change of the friction law from a mesoscopic level to a macroscopic
level is studied in the spring-block models introduced by
Burridge-Knopoff. We find that the Coulomb law is always scale invariant.
Other proposed scaling laws are only invariant under certain conditions.}
\vfill\eject
The first documented
studies on friction were done by Leonardo da Vinci, who experimentally
verified two basic laws of friction.  His studies were rediscovered by
Amonton de la Hire who announced the two laws in 1699 in the following form (a)
the friction force is independent
of the size of the surfaces in contact, and (b) friction is proportional
to the normal load.
Nearly 100 years after Amonton, Coulomb recognized the difference between
static and dynamic friction. He noticed that the initial friction
increased with the time the surfaces were left in stationary contact.
During the slip process the friction is smaller than the static
friction, and he proposed  the dynamic friction to be constant (For these
accounts see for example Cap. 2 of Ref. [1]).

More recent experimental studies on friction show, however, that
friction does vary during sliding, decreasing with slip[2,3].
This is called {\sl slip weakening}. In particular, if the friction
has a negative dependence on the sliding velocity it is
called {\sl velocity weakening}.
In these cases a dynamic instability
can occur, resulting in very sudden skip with an associated stress
drop. This often occurs repetitively -  a period of rest follows
the slip, which in turn is followed by a period of rest, and so on.
This behavior is called stick slip motion.
It is commonly observed in the frictional
sliding of rocks, which led Brace and Byerlee[4] to propose
it as a mechanism for earthquakes.
Besides the stick slip process,
two surfaces with friction can also have a steady relative motion, with the
sliding
velocity  constantly positive.

In an experimental investigation one understands  friction force as
a macroscopic average of the resistance
to motion due to
microscopic interactions between two sliding surfaces.
The microscopic forces have
various sources and they are a topic of great interest nowadays.
Several theoretical models for friction have been proposed
recently[5,6].
In these models one considers a mesoscopic level and studies how
individual elements (having a given mesoscopic friction behavior) synchronize
to a collective behavior giving the macroscopic phenomena like
the classical Coulomb law or the stick slip motion.
The question arises -
Will these mesoscopic friction laws reproduce the macroscopic behavior observed
experimentally,
and are they consistent with the phenomenological models that have
been proposed? This fundamental question is the subject of this paper.

We investigate two mechanical models
introduced by Burridge and Knopoff[7] to
mimic the dynamics of earthquakes. Each model consists of a chain of blocks
connected by springs, the set being driven at constant velocity on a surface
with
friction.  Several studies of these models have been performed showing
that, at least qualitatively, that they can present the dynamics observed in
real earthquakes [7-10].
We study the relation between the macroscopic and mesoscopic behavior of
friction
using four different friction laws, which are introduced here in the
same chronological order they appeared in the literature.

The first model is
shown schematically in Fig. 1(a).
It consists of
a chain of $N$ blocks of mass $m$ coupled to each other by harmonic
springs of strength $k$. The blocks are on a surface and
the first one is pulled with constant
velocity $v$.
Between the blocks and the surface acts a velocity dependent
frictional force $f$, which is usually nonlinear, giving the system
a rich behavior.
This model has been called ``the train model"[9] to
distinguish it from the other model, which was also introduced
by Burridge and Knopoff.

The equation of motion for a moving block $j$ is given by
$$m\ddot X_j=k(X_{j+1}-2X_j+X_{j-1})-f(\dot X_j/v_c), \ \ \ \dot X_j\neq
0,\eqno(1)$$
with $j=1,\dots, N$.
$X_j$ denotes
the displacement of the $j$-th block measured
with respect to the position where the
sum of the elastic
forces in the block is zero.
The open boundaries are given by $X_{N+1}=X_N$ and $X_0=vt$.
In Eq. (1) we consider the friction force as
function of the instantaneous
block velocity with respect to a characteristic velocity $v_c$.
If we write
$f(\dot X/v_c)= f_{\circ}\Phi(\dot X/v_c) $
where $\Phi (0)=1$
and introduce the variables
$\tau =\omega _p t,\ \ \omega _p^2=k/m,\ \ U_j=kX_j/f_{\circ}$,
Eq. (1) can be written in the following dimensionless form[9]
$$\ddot U_j=U_{j+1} -2U_j+U_{j-1}-\Phi(\dot U_j/\nu_c),\ \ \ \dot U_j\neq 0,
\eqno (2)$$
with $U_{N+1}=U_N$, $U_0=\nu \tau$,
$\nu=v/V_{\circ}$,
$\nu_c=v_c/V_{\circ}$, and $V_{\circ}=f_{\circ}/\sqrt{km}$.
Dots now denote differentiation with respect to $\tau $. In a system
of a single block the quantity $f_{\circ}/\omega _p$ is the
maximum displacement of the pulling spring before the block
starts to move; in the absence of dynamical friction
$2\pi/\omega _p$  and $V_{\circ}$ are respectively
a characteristic period of oscillation of the block and the maximum velocity it
attains.
This system is completely described by  two dimensionless parameters, $\nu $
and $\nu _c$.

The other model we study is shown schematically in Fig. 1(b).
It consists of $N$ identical blocks of mass
$m$ connected to each other through linear springs of constant
$k_c$. The system is driven by an upper bar moving at constant
velocity $v$ with respect to the lower supporting surface at rest.
Each mass is attached to the upper bar by a linear spring of
constant $k_p$. Between the masses and the supporting surface there is
again a
velocity dependent friction force $f$. We will call this system the BK model.

The equation of motion for a moving block $j$ is given by
$$m\ddot X_j=k_c(X_{j+1}-2X_j+X_{j-1})-k_p(X_j-vt)-f(\dot X_j/v_c), \ \ \ \dot
X_j\neq 0,\eqno(3)$$
with $j=1,\dots, N$.
We use open boundary conditions, which are given by $X_0=X_1$ and
$X_{N+1}=X_N$.
$X_j$ denotes
the displacement of the $j$-th block measured from the point
where the sum of all the elastic
forces in the block is zero.

Normalizing the
parameters and variables in a similar way as above (where $k$ is replaced
by $k_p$)
the BK model
has the following equation of motion[8]
$$\ddot U_j=l^2(U_{j+1} -2U_j+U_{j-1})-U_j+\nu \tau-\Phi(\dot U_j/\nu_c), \ \ \
\dot U_j \neq 0,\eqno (4)$$
with $l^2=k_c/k_p$.
$l$ can be interpreted as the velocity of the sound in the chain.
The system has therefore three fundamental parameters, namely, $l$, $\nu$ and
$\nu _c$.

The friction force $\Phi$  models the mesoscopic interaction
between the blocks and the surface. We call it the
{\sl mesoscopic } friction force. The {\sl macroscopic} friction force,
denoted here by $\overline F$, is given by
a temporal average of the force applied by the driving
mechanism normalized to the number of blocks.  For the train and the BK models
the applied forces are given
respectively by
$$F_t=\nu \tau- U_1 , \eqno(5) $$
$$F_{BK}= \nu \tau - \sum_{j=1,N} U_j. \eqno(6) $$
Thus, $\overline F= <F>/N$, where $<...>$ stands for temporal average.
The normalized friction force is in fact what is called
the friction coefficient.

Unless otherwise stated, in the numerical simulations
we start the system
with all the blocks
at rest. In the train model the initial position for each block is
taken $U_j=0$, i.e., at equal distances.
Even with a perfectly homogeneous initial configurations a complex dynamics may
naturally emerge in the train model[9].
In the BK model a small inhomogeneity in the positions of the
blocks must be introduced, otherwise the system will not present
any complex behavior, and the chain will move as a single block[8].
For computational convenience we do not allow backward motion of
the blocks, that is, the friction force will attain a sufficiently
high value in order
to forbid backward motion. Tests have been performed where backward motion
was allowed and we did not observe any significant difference from
the results shown here.
In our calculations we use open boundary conditions for the chain. Several
tests have shown that our results do not change if  periodic
boundary conditions are used.

We have studied for $\Phi $ four different functions found in literature to
verify which
ones will conserve their functional form when going from the mesoscopic
to the
macroscopic level in the mechanical models introduced by Burridge and
Knopoff. Since scale invariance is often observed in the roughness of
solid surfaces, a functional similarity of the friction law seems to
be a condition that should be fulfilled in order that this law can
be applied in these cases.
\bigskip
\noindent {\bf (a) Coulomb force}
\medskip
The friction law formulated by Coulomb is independent of the value of the
sliding velocity  and
is a discontinuous function at zero velocity given by
$$\Phi (\dot U/\nu _c)=\cases{1, &if  $\dot U=0$;\cr
\Phi_{\circ}, &if $\dot U>0$, \cr}\eqno(7)$$
where $\Po < 1$.
Note that using this friction force the only nonlinearity in the system
is  this discontinuity in $\Phi $.

To clarify the dynamics of the mechanical models using Coulomb
force,  we start by investigating the dynamical evolution of a one-block
system.
The equation of motion for this system in dimensionless quantities
is
$$\ddot U=-U+\nu \tau -\Po, \ \ \ \ \dot U>0. \eqno (8)$$
A possible solution for this equation is
$U^e=\nu \tau -\Po$, which gives $\dot U^e= \nu $.
That is, the block
is moving with constant velocity, equal to the pulling speed.
The superscript $e$
denotes equilibrium position.
The stability of this solution can by investigated by perturbing it. If one
writes
$U=U^e+u\exp(\omega \tau)$ and substitutes this equation into Eq. (8),
one will find  $|\omega| =1$, which implies
that the solution with the block having constant velocity
is marginally stable.
A nontrivial  solution  is
$$U=(\nu _o-\nu)\sin \tau +(U_o+\Phi_{\circ})\cos \tau+ \nu \tau-\Phi_{\circ},
\eqno(9)$$
where $U_o$ and $\nu _o$ denote $U(\tau =0)$ and $\dot U(\tau=0)$,
respectively. If $\dot U=0$  and the elastic force is smaller than
the static friction, the block will stick to the surface until the
elastic force becomes  again larger than the static frictional force.
The evolution of the system in in phase space is shown  in Fig. 2(a)
for a given example.
It consists of concentric ellipses around the trivial
solution ($U=U^e$ and $\dot U=\dot U^e$)
when the initial conditions are such that
$(\nu _o-\nu)^2+(U_o+\Po)^2<\nu ^2$. If this condition is not satisfied,
then the block will stick to the surface, and  harmonic motion
will occur only in the upper region of the phase space, where $\dot U > 0$.
The trajectories
in phase space evolve symmetrically around the
trivial solution,
which allows us to conclude that the average elastic force is
given by $\overline F= \Po$ in both cases, that is, when the motion
is steady or via stick slip.

For the train model with $N>1$, the trivial situation in which all
the blocks move with constant velocity,  equal to the
pulling velocity, has  $\ddot U_j=0$ for each $j$.
{}From Eq. (2) we find
$$U^e_{j+1}-2U^e_j+U^e_{j-1}=\Po, \ \ \ \ j=1,...,N \eqno(10)$$
with $U^e_0=\nu \tau$ and $U^e_{N+1}=U^e_N$. The superscript $e$ in $U$ denotes
again
equilibrium position.
For the first block, $j=1$, the equilibrium position is
easily found. If one adds the equations given by Eq. (10) with $j=1,...,N$,
one will find  $U^e_1=\nu \tau -N\Po$. This gives $\overline F_t=\Po$.
The equilibrium position
for any other block
is a function of the system size, and is found by solving the
system of equations given by Eq. (10).

In situations where no block of the chain sticks to the surface (this occurs
for example when $\Po=1$),
the motion of the first block is governed by
$$\ddot U_1=-\Omega ^2 U_1 +\nu \tau/N-\Phi_{\circ}, \eqno(11)$$
where
$\Omega ^2=1/N$.
The solution of Eq. (11) is
$$U_1={{1}\over {\Omega }} (\nu _o-\nu ) \sin (\Omega \tau)+(U_o + N\Phi
_{\circ })\cos (\Omega \tau)+\nu \tau- N\Po, \eqno(12)$$
where $U_o$ and $\nu _o$ are the position and velocity of the first block
at $\tau =0$.
The motion of the first block is harmonic and it occurs in phase space
around
the equilibrium solution  $U_1= U^e_1$ and $\dot U_1=\dot U^e_1=\nu $.
The average value of $U_1$ obtained from Eq. (12) is
$\overline U_1= \nu \tau  - N\Po $.
If
we use this result in Eq. (5) we find $\overline F_t= \Po  $.

When $\Po < 1$ the
blocks can in principle stick to the surface. In this case, the evolution of
the
system becomes nonlinear, and the integration of the equations of
motion does not seem to be very simple as for a one-block
system. However, even when sticking
occurs in a system with more than one block
we find numerically that $\overline F_t=\Po$. In all investigated cases
we found that the first block always has a symmetric trajectory around
the trivial solution, whether or not blocks stick to the surface.
As a simple example, we discuss a two-block system ($N=2$).
Take  $\Po =0.8$ and
the initial positions and velocities given respectively by
$ U_1=U_2=0$, $\dot U_1=\dot U_2=0$. If $\nu =0.1$ we see that
the first block sticks to the surface and
executes a period two orbit around the trivial solution
$U^e_1=\nu \tau - 2\Po$. If the pulling velocity
is increased to $\nu =1$, the motion of the first block becomes
more complicated, and executes what seems to be a quasiperiodic
orbit, but also symmetric around the trivial solution.
These two different motions are shown in Fig. 2(b) and 2(c),
respectively.
For larger systems the orbits described by the blocks
in phase space become more and complicated. However, the
first block seems always to have
a symmetric trajectory around the
trivial solution, which from Eq. (5) results in $\overline F_t=\Po $.

For the BK model, shown in Fig. 1(b),
the macroscopic force
can also be written as
$$F_{BK}=\nu \tau -NW,\eqno (13)$$
where $W$ is the coordinate of the center-of-mass, defined as
$$W={{1}\over{n}}\sum _{j=1,N} U_j. \eqno (14)$$
Thus, for this model is the evolution of $W$, and not of $U_1$, that will
determine the
the value of the macroscopic force.

The evolution for $W$ when  no block sticks to the surface
is found by adding Eq. (4)
with $j=1,...,N$. The following expression  is obtained
$$\ddot W=-W +\nu \tau -\Phi_{\circ}. \eqno (15)$$

The trivial solution in which all blocks move with the same velocity has
$\ddot U_j=0$,  for all $j$.  This results in $W^e= \nu \tau -\Po$. Using
this result in Eq. (14) we find
for this trivial motion  $\overline F_{BK}=\Po$.
A simple linear stability analysis shows that this solution is marginally
 stable.
A nontrivial solution for Eq. (15) is
$$W=(\nu _o-\nu)\sin \tau +(W_o+\Phi_{\circ})\cos \tau+\nu \tau -\Phi_{\circ},
\eqno(16)$$
where $W_o$ and $\nu _o$ denote $W(\tau =0)$ and $\dot W(\tau=0)$.
The motion for $W$ is harmonic and it evolves around
$W=W^e$ and $\dot W=\dot W^e$, which gives again $\overline F_{BK}=\Po$.

These results show that when sticking does not occur
(this is the case,  for example, when $\Po=1$)
the motion of
center-of-mass of the chain has an equation completely identical
to the equation of the one-block system (Eq. (8)).
When the nonlinearity
in $\Phi $ is present we have noticed that the coupling between the
blocks strongly affects the dynamics of the system. To exemplify
this we show the temporal evolution of a BK system
with and without coupling when it is governed by the Coulomb law.
Fig. 3 consists of diagrams of the block number $j$ versus the time $\tau$
with a dot representing $\dot U_j>0$. For Fig. 3(a)
the coupling between
the blocks is zero, that is, $l=0$. The starting conditions are such that $\dot
U_j(\tau=0)=0$ and
$U_j(\tau=0)$ is randomly distributed between $[-0.01,0.01]$. The parameters
are $\Po=0.8$, $\nu =0.01$ and $N=100$. We see that the blocks periodically
stick
to the surface, as also seen in Fig. 2(a). When the coupling is added
between the
blocks the behavior becomes very different, as Fig. 3(b) shows.
There we have taken
$l^2 =50$. We see pulses traveling through the chain with the
sound velocity, and the white regions have smaller area than
in the case where the coupling is zero.
A transient time has been discarded in all simulations shown in this paper.
For the train model we start our computations after the last block has moved.
For the BK model we discard the time corresponding to ten loading periods,
where
the loading period is given by $1/\nu $.

We have investigated  numerically the
evolution of $W$ when sticking
to the surface
is in principle possible. We observed
that a plot of $\dot W$ versus $W-\nu \tau$ is qualitative similar
to trajectories shown in Fig. 2(a).
This results in $\overline F_{BK}=\Po$.

We have also performed numerical simulations in larger chains.
In figure 4(a) and 4(b) we show  temporal evolutions for  $F_t$ and $F_{BK}$
using Coulomb law, with $N=200$ for the train
model and $N=500$ for the BK model. The other parameter values are
$\nu=5$, $\Po=0.8$ and $l^2=50$. In Fig. 4(a) we see that the
evolution of $U_1$, although complex,  seems to be periodic.
For the
BK model, the evolution of $F$ is also periodic, and
has a much simpler structure.
The frequency of $F_{BK}$ is one, which is the natural frequency
associated with the center-of-mass motion, as found from
equation (15).
In both cases we find $\overline F=\Po$.

We have varied the pulling velocity $\nu $ and the numerical values found for
the macroscopic force in the two models
are shown in Fig. 4(c).  The parameters used are the
same as in Figs. 4(a) and 4(b), with exception of $\nu $, which now is
varying.
We represent the mesoscopic friction force $\Phi$
by the solid line, the macroscopic force $\overline F$ by circles for the train
model and
by squares for the BK model.
There is an excellent agreement
between the circles and the squares with the solid line.
The small deviations are within the statistical errors.
Therefore, we have observed that the Coulomb law gives the same behavior for
the mesoscopic force as for  the macroscopic force in both models.
In other words, our results show that the Coulomb friction force is
scaling-invariant and the procedure can be iterated in a renormalization
approach.
\bigskip
\noindent {\bf (b) Dietrich-Scholz friction force}
\medskip
In experimental studies on rock friction, Dietrich[2] and
Scholz {\sl et al.}[3] have found that the friction force  has
a logarithmic dependence on the sliding velocity, decreasing as
the velocity increases.
They proposed a friction force of the form
$$\Phi(\dot U/\nu _c) =\Phi_{\circ}-b \log (\dot U/\nu_c). \eqno(17)$$
It is clear that, due to the logarithmic dependence, cutoffs have to be
introduced to this function for high and low
velocities.
We introduce cutoffs for small and large  velocities in such a way that
the function $\Phi $  remains continuous. We consider, then
$$\Phi (\dot U/\nu_c)=\cases{1, &if $ \dot U < \nu_c$; \cr
              1-b\log (\dot U /\nu_c) , &if $\nu_c \le \dot U \le \nu_c\exp
(1/b)$; \cr
              0 , &if $ \dot U> \nu_c\exp (1/b)$. \cr} \eqno (18)$$
For small pulling velocities, in both models, the chain experiences only the
linear part of the friction force with $\dot U \le \nu _c$, where we
have $\Phi_{\circ}=1$.
In this case, stick of blocks to the surface does not occur and analytical
results can be found for $\ft $ and $\fbk$, as shown
in the previous subsection.

The nonlinear regime is attained when the maximum velocity
attained by a block is equal or greater than $\nu _c$.
For the train model we have to consider two
distinct situations. When we start the system with all the blocks
at rest, a simple linear analysis, similar to the one performed in
subsection (a), shows that the maximum velocity attained by the
blocks is equal to $\nu $, if the
slipping event does not involve all the blocks of the system. If all the blocks
are involved
in the event, then the maximum velocity of the blocks is
$2\nu $. This can easily be found
from Eq. (12).
Thus, when $\nu < \nu _c/2$ the chain never feels the the nonlinear
part of Eq. (18) and moves continuously.
This results in $\overline F_t=\Po =1 $.
When $\nu _c/2 \le \nu \le \nu _c$ the  events that do not involve all
the blocks of the chain
will not feel the nonlinear regime, whereas the events where all
the blocks of the chain are displaced will be in the nonlinear regime
of the friction force.
In this situation, as time evolves $F_t$ increases monotonically until the
last block moves, then we see a sharp drop in $ F_t$. Then $F_t$ starts
to increase monotonically again,  and the process repeats.
For $\nu > \nu_c$ the motion, in any event, will be
completely in the nonlinear regime and stick slip dynamics may
be observed.
Now a complex evolution is seen in $F_t$. As an example,
in Fig. 5(a) we show the temporal evolution for $F_t$
where we use the parameters $N=200$, $\nu _c=1$,
and $\nu =5$. Here $F_t$ does not seem
to be globally periodic.

For the BK model, when only the linear part of the
friction force is felt,
the chain behaves
as a single block, as  discussed in subsection (a).
The motion for the center-of-mass is
governed by Eq. (16) with $\Phi_{\circ}=1$.
{}From there we get
$\dot W=\nu [1-\cos \tau]$.
Consequently, the
maximum velocity attained by the blocks is given by $\dot W_{max}=2\nu$. From
here we find that when $\nu \ge \nu_c/2$ the nonlinear part of
the friction force given by Eq. (18) will be felt, and the motion
can be via stick slip.

We show in Fig. 5(b) an example for the temporal evolution of $F_{BK}$.
The parameters are $N=500$, $l^2=50$, $\nu _c=1$, $b =0.3$ and $\nu =5$.
Since $\nu > \nu_c$ the motion is nonlinear.
We see a roughly  periodic function with variable amplitude. The
frequency here is one,
which is the natural frequency of the center-of-mass motion, as
given by Eq. (15).

We plot in Fig. 5(c) the mesoscopic friction force $\Phi $ (solid line)
and the macroscopic forces $\overline F$ for the train model (circles) and for
the
BK model (squares) as  functions of the
pulling velocity. The respective
parameters are the same as in Fig. 5(a) and 5(b).
For the BK model there is a good agreement between the
mesoscopic and macroscopic force.
By rigidly shifting the mesoscopic force by a given factor, we will be
able to make the curves practically coincide in the logarithmic region.
The curves also coincide at the plateaux,
where we have $\Phi _{\circ}=1$ and $\Phi _{\circ}=0$, as expected.
But in this case no shift of the mesoscopic force is necessary.
We have studied other regions of the parameter space and  also found
a good agreement between
$\overline F_{BK}$ and $\Po $.

The comparison between the macroscopic and mesoscopic force for the
train model with this  friction force shows,
however, that the two curves do not present the same behavior in the
logarithmic regime.
This is clearly seen in Fig. 5(c) where the results for this model are
represented by circles.
In the region where  the dynamics is linear
the numerical results agree with the analytical ones. At $\nu = \nu _c/2$
we see the expected transitions to the nonlinear regime.
When the motion becomes nonlinear
there is a  jump in
$\overline F_t$
and the macroscopic force
becomes
much smaller than the mesoscopic one.

So, for the friction force proposed by Dietrich and Scholz
we find a good agreement between the macroscopic and mesoscopic
friction forces only for the  BK model, so that the
friction law seems to be scale invariant. For the
train model there is no  agreement between the two forces.
\bigskip
\noindent {\bf (c) Carlson-Langer friction force}
\medskip
Carlson and Langer[8] have studied the dynamics of the BK
model using the velocity weakening friction force
given by
$$\Phi (\dot U/\nu _c)=\cases{1, &if $\dot U=0$; \cr
        [1+\dot U/\nu _c]^{-1},  &if $\dot U>0$. \cr} \eqno(19)$$
The train model with the friction force given by Eq. (19) was investigated
in Ref. [9]. It was found that it presents slipping
events of several sizes, and the distribution of event sizes was
studied in detail.

The dynamics of the BK model, when governed by a friction force
given by Eq. (19), has been investigated extensively in recent publications[8,
10, 11].
It has been shown that when $\nu $ is small there are basically two kinds of
slipping events in the
chain, which are the small and the large events. The small events displace a
small number of blocks and they
never attain velocities greater than $\nu _c$.
The small events, although numerous, relax almost
no elastic energy, and they disappear in the continuum limit[11].
The small events also disappear
for large $\nu $.
The big events displace a large number of blocks
($n \approx 4l^2\nu _c \ln [4 l^2/\nu ]$, as found in [8])
and they relax almost the total stress accumulated in the chain.
They occur in a nearly periodic way.

In Figs. 6(a) and 6(b) we see  temporal evolutions  of the macroscopic
force for train and BK models, respectively. We have
taken $\nu _c=1$ and $\nu =1$. For the train model
we see fluctuations of several sizes, with the largest ones representing
the situation where all blocks of the chain are displaced.
For the BK model the periodic character of the center-of-mass motion is
also apparent here. The frequency is one. In these
simulations
we have used $N=200$ for the train model, $N=500$ and $l^2=50$  for the BK
model.

The macroscopic force as a function of the pulling speed is shown
in Fig. 6(c). We see that for the train model (circles) $\overline F_t$ is much
 smaller than
the mesoscopic force (solid line) .
We have studied  other regions of the parameters space and no fixed line was
found for the macroscopic and mesoscopic behavior in the train model
when this friction force is used.

For the BK model (squares) we see a very good
agreement between the mesoscopic and macroscopic forces.
Therefore, for these parameter
values, the macroscopic
behavior is consistent with the mesoscopic one. The results shown
are insensitive to variations in $l^2$, but present
sensitivity on $\nu_c$ for small pulling speeds ($\nu \ltsim \nu_c$).
If we decrease the characteristic velocity, that is, for $\nu _c < 1$ we
see that the macroscopic force is much smaller than the mesoscopic
one for small velocities, whereas  the macroscopic force
is significantly larger than the mesoscopic force if $\nu _c >1$.
For larger pulling velocities ($\nu \gtsim \nu_c$) we always find
good agreement between mesoscopic and macroscopic forces.

It has been shown [10] that the BK model presents a dynamic phase transition
when $\nu _c = 1$. For characteristic velocities greater than this
critical value the system moves in a continuous way. When $\nu _c < 1$
the motion is via stick slip.
Thus, we see an excellent  agreement between the macroscopic and
mesoscopic forces (for all values of $\nu$)  only at the critical point
$\nu _c=1$, where the motion
changes its character. When the stick slip motion occurs ($\nu _c < 1$), much
less force is necessary to displace the chain and one has a significant
discrepancy between
the macroscopic and mesoscopic friction forces.
So, the stick slip motion seems to be an intelligent
way the system uses to move with less effort.

In summary, the friction force suggested by Carlson and Langer
has the same behavior in the macroscopic and mesoscopic levels only
for the BK model at the critical velocity $\nu _c=1$.
A self-consistent relation between the macroscopic
and mesoscopic forces is not found in the train model.
\bigskip
\noindent {\bf (d)  Schmittbuhl-Vilotte-Roux force}
\medskip
Schmittbuhl {\sl et al.} [6] recently investigated the average
friction force on a rigid block
that slides
on a self-affine
surface. Taking into consideration that the block  jumps
on ballistic trajectories, they found that for large velocities the
friction force appears to be proportional to $v^{-2}$. For low
velocities they found that the apparent friction coefficient is constant.
The crossover between the two regimes is determined by
a characteristic velocity. The equation we consider for this friction force
is given by
$$\Phi(\dot U/\nu _c)=\cases{1, &if $\dot U<\nu_c$; \cr
              {(\dot U/\nu _c)^{-2}} , &if $ \dot U \ge \nu _c$, \cr}
\eqno(20)$$
where we have introduced a cutoff at $\dot U=\nu _c$.
When the chain experiences only the linear part of the friction force,
that is, when
no block attains velocity equal or greater than $\nu _c$,
an analytic solution is for the motion of the center-of-mass is  possible
for both models, as we
have seen in subsection (b). There we have shown that
the nonlinear part of the friction force is felt when $\nu \ge \nu _c/2$.
In the train model there is an intermediate regime,
which occurs when $\nu _c/2 \le \nu \le \nu _c$.
In this regime $F_t$ increases monotonically until the last block
of the chain is displaced. Then, a sharp drop in $F_t$ seen. After this, the
macroscopic force starts to increase again, and the process repeats.
For
$\nu > \nu _c$ the motion
of the train model
becomes fully nonlinear and stick slip behavior can be observed.

We show a temporal
evolution of $ F_t$  with $N=200$, $\nu _c=1$ and
$\nu =1.5$ in Fig. 7(a) and of $F_{BK}$ in Fig. 7(b)
with $N=500$, $l^2=50$, $\nu _c=1$ and $\nu =1.5$.
Again, we see a complex evolution for
$F$ in the train model with the largest jumps representing displacement of all
blocks.
For the BK
model
$F$ presents a nearly periodic behavior with variable
amplitude and frequency roughly equal to one.

The results of our simulations for the dependence of $\overline F$ with $\nu $
are shown in Fig. 7(c).
For the train model (circles) we see a sharp drop in the macroscopic friction
force
when the motion becomes nonlinear, i.e.,
when $\nu \ge \nu _c/2$. For this model
we find that asymptotically (for large $\nu $)
the macroscopic and mesoscopic (solid line) forces seem to present the same
behavior. It is difficult the study of $\overline F_t$ for much larger
velocities as the ones we investigated because the statistical errors get
very large and become of the order
of $\overline F_t$.

For the BK model we see that the
macroscopic force (circles) coincides with the mesoscopic one only in
the plateaux region,
where $\Phi =1$, and the dynamics is fully linear.
There is also a single point where the curves cross each other.
We conclude  that the friction force suggested
in Ref. [6] does not give the same behavior for the macroscopic
and mesoscopic levels in the BK  model.
However, note that in this case the macroscopic force has also a power
law dependence on $\nu $, but the exponent is
clearly different from the exponent of the mesoscopic force.

In conclusion, we have studied the scale invariance of various proposed
friction
laws on two spring-block models of Burridge and Knopoff.
We find that  the classical
Coulomb law, namely no velocity dependence of the friction force for
finite velocity, is always invariant under a change from the mesoscopic to
the macroscopic scale. The friction law proposed by Dietrich and Scholz
are only invariant if the model with an upper bar is considered.
The friction law used by Carlson and Langer is only invariant at the
critical velocity. The relation recently suggested by Schmittbuhl {\sl et al.}
seems to be scale invariant for large pulling velocities in the train model.
However,
more precise simulations would be required for a definitive assessment.

Since the self-affinity of rough surfaces has been extensively documented,
in particular for rocks the scale invariance seems to be an important
condition for the force law. It would, in fact, be better to formulate
a renormalization procedure for which the scale-invariant law would
be ``fixed points". This might allow to make a direct connection between
the scaling of the friction force in terms of the roughness exponent.
Finally, would it be also important to go to the microscopic scale
by considering the geometrical hindrances represented by the asperities on
rough surfaces.

\medskip
\bigskip
\noindent{ACKNOWLEDGMENTS}
\medskip
\medskip
We thank S. Roux and J. P. Vilotte for many enlightening discussions.
MSC thanks the Alexander von Humboldt fundation for financial support, the
hospitality at
Hochstleistungsrechenzentrum and at Universit\"at Essen,
where this work was started.
\bigskip
\medskip
\noindent {REFERENCES}
\bigskip
\item {1.} C. H. Scholz, {\sl The Mechanics of Earthquakes and
Fauting}, Cambridge Univ. Press, New York, 1990.

\item {2.} J. Dietrich, {\sl Pure Appl. Geophys.} {\bf 116}, 790 (1978).

\item {3.} C. H. Scholz and T. Engelder, {\sl Int. J. Rock Mech. Min. Sci.}
{\bf 13}, 149 (1972).

\item {4.} W. F. Brace and J. D. Byerlee, {\sl Science} {\bf 153}, 990 (1966).

\item {5.} T. P\" oschel and H. Herrmann, {\sl Physica A} {\bf 198}, 441
(1993).

\item {6.} J. Schmittbuhl, J-P. Vilotte and S. Roux, to appear in {\sl J. de
Physique}.

\item {7.} R. Burridge and L. Knopoff, {\sl Bull. Seismol. Soc. Am.} {\bf 57},
341 (1967).

\item {8.} J. M. Carlson and J. S. Langer, {\sl Phys. Rev. Lett.} {\bf 62},
2632 (1989); {\sl Phys. Rev. A} {\bf 40}, 6470 (1989).

\item {9.} M. de Sousa Vieira, {\sl Phys. Rev. A} {\bf 46}, 6288 (1992).

\item {10.} G. Vasconcelos, M. de Sousa Vieira and S. R. Nagel, {\sl Physica
A} {\bf 191}, 69 (1992); M. de Sousa Vieira, G. Vasconcelos and S. R. Nagel,
{\sl Phys. Rev. E} {\bf 47}, R2221 (1993).

\item {11.} J. Schmittbuhl, J-P. Vilotte and S. Roux,
{\sl Europhys. Lett.} {\bf 21}, 375 (1993).

\vfill\eject
\noindent{FIGURE CAPTIONS}
\medskip
\medskip
\item {Fig. 1.} (a) Train model,  which consists of a chain of blocks
connected by linear springs. The blocks are on a flat surface and the
first one is pulled with constant velocity. (b) BK model where each
block is connected to a driving bar and to the neighboring blocks.

\item {Fig. 2.} (a) Evolution in phase space of a one-block system for
different initial conditions. The central dot represents the trivial
solution in which the block has constant velocity equal to the pulling
speed $\nu =0.1$. The ellipses have initial conditions given by
$U(\tau=0)=-\Po=-0.8$ and
$\dot U(\tau=0)=0.05,\  0.025,\  0.01,\  0$.
The initial velocities decrease in the outward direction.
(b) Evolution in phase space of the first block in a
two-block train system with $\Po =0.8$ for $\nu =0.1$ giving a
period two orbit.
(c) The same as (b) with $\nu =1$, which seems to result in a quasi-periodic
motion.

\item {Fig. 3} (a) Temporal evolution of the BK model with $N=100$, $\Po=0.8$,
$\nu =0.01$ with (a) $l=0$ and (b) $l^2=50$.
The diagrams show the block number $j$ versus $\tau $.
A dot in the figures means $\dot U_j > 0$.

\item {Fig. 4.} (a) Temporal evolution of the applied
force $F_t$ for the train model  and (b) of $F_{BK}$ associated
with the
BK model. The mesoscopic force is
given by the Coulomb law with $\Po =0.8$. The parameters are
$N=200$ and $\nu =5$
for the train model and
$N=500$, $l^2=50$, $\nu =5$ for the BK model.
(c) Mesoscopic force $\Po $ (solid line) and macroscopic
forces $\overline F$  versus the pulling velocity for
the train model (circles) and the BK model
(squares).
With exception of $\nu $, the parameter values are the same as in (a).
The average macroscopic force was calculated in the train model for
1,000,000 time steps using intervals of $\delta \tau= 0.02$.
For the BK model
the temporal average was calculated for 200,000 iterations using
time intervals of $\delta \tau=0.01$.

\item {Fig. 5.} (a), (b) and (c),
the same as in Fig. 4(a), 4(b) and 4(c),  respectively,
for the friction force given
by Eq. (18), with $\nu _c=1$, $b=0.3$, and with $\nu =5$ in (a) and (b).
In Fig. 5(c) and 7(c) we needed to increase the number of
iteration steps (in relation to Fig. 4(a)) in order to decrease the statistical
erros associated
with large pulling velocities.

\item {Fig. 6.} (a), (b), and (c),  the same as in Fig. 4(a), 4(b), and
4(c), respectively,
for the friction force given
by Eq. (19), with $\nu _c=1$ and with $\nu =1$ in (a) and (b).
The mesoscopic force here has been rigidly shifted  by a constant
factor in order to make it coincide with the macroscopic force of the BK
model.

\item {Fig. 7.} (a), (b) and (c), the same as in Fig. 4(a), 4(b) and
4(c), respectively,
for the friction force given
by Eq. (20), with $\nu _c=1$ and with $\nu =1.5$ in (a) and (b).

\end